\documentclass[epj]{svjour}
\usepackage{graphics}
\begin{document}
\title{Deuteron Photodisintegration in the Quark-Hadron Picture}
\author{Patrizia Rossi for the CLAS Collaboration
}                     
\institute{INFN - Laboratori Nazionali di Frascati - Via Enrico Fermi 40, 00044 Frascati (Italy)}
\date{Received: date / Revised version: date}
%
\abstract{
The study of the two-body photodisintegration of the deuteron in the few GeV region is the ideal reaction to clarify the transition from the nucleonic degrees of freedom to the QCD picture of hadrons. The CLAS large angle spectrometer of Hall B at Jlab allowed for the first time the complete measurement of the angular distribution of the differential cross section at photon energy between 0.5 and 3 GeV. Preliminary results of the E93-017 experiment from the analysis of the 30\% of the total statistic accumulated show persistent forward-backward asymmetry and are well described by the recent calculation of the deuteron photodisintegration cross section derived in the framework of the Quark Gluon String Model.
\PACS{
      {PACS-key}{discribing text of that key}   \and
      {PACS-key}{discribing text of that key}
     } 
} 
\maketitle
\section{Introduction}
\label{intro}
One of the major goal of Jlab is to explore and study the interplay between the nucleonic and partonic pictures of the strong interaction. Although the standard nuclear models are successful in reproducing the overall picture of hadrons interacting at large distances and QCD is convincing in the description of quarks interacting weakly at small distances (pQCD), the physics connecting the two regimes is almost non existent. When probing distances comparable to those separating the quarks, classical nuclear physics should break down at some point, yet the nucleonic picture still describes many features of the strong interaction. The alternative is to look for the onset of experimentally accessible phenomena which are naturally predicted by QCD.\\
Deuteron photodisintegration is well suited to address the intermediate energy region in order to clarify the transition from the hadronic to the partonic description of hadrons and nuclei. This because: {\it i)} large momentum transfers to the constituents can be obtained with photon energies of only a few GeV \cite{holt}, {\it ii)} the deuteron is the simplest nucleus (A=2) and {\it iii)} the electromagnetic probe is well known.\\
The first measurement of the $\gamma d \rightarrow p n $ differential cross section in the GeV region and for proton angle $\theta_{CM}= 90^{\circ}$ \cite{SLACNE81} showed the onset of asymptotic scaling near a photon energy of 1 GeV. This result stimulated a renewed interest in studying this reaction: experiments were planned to continue at SLAC \cite{SLACNE82},\cite{SLACNE17} and a broad program of physics, including the measurement of the recoil proton polarization, has been started at Jlab in all the three experimental halls \cite{Bochna},\cite{Schulte},\cite{Gilman1},\cite{Gilman2},\cite{Rossi}.
The situation is summerized in Table 1. It is worth noticing that the experiment E93-017 \cite{Rossi} is the only one covering a very broad angular and energy range.
%
%
\begin{table}
\caption{$\gamma d \rightarrow p n $ experiments in the last 15 years }
\label{tab:1}       
\begin{tabular}{lll}
\hline\noalign{\smallskip}
$\theta_{CM}$(deg) &$E_{\gamma}$ (GeV) & Ref  \\
\noalign{\smallskip}\hline\noalign{\smallskip}
90   & 0.8 1.1 1.3 1.6 & \cite{SLACNE81} \\
52 66 78 90 113 126 142   & 0.8 1.0 1.2  & \cite{SLACNE82} \\
90 113 142   & 1.4 1.6 1.8 &  \cite{SLACNE82} \\
37 53 89       & 1.5 1.9 2.3 2.7 & \cite{SLACNE17}\\
36 52 69 89    & 0.8 1.5 2.4 3.2 4.0 &  \cite{Bochna} \\
37 53 70       & 5.0 5.5 &  \cite{Schulte} \\
90 (proton polarization)             & $0.8\div2.4$ &  \cite{Gilman1} \\
30 36 52 70 90 110 127 142    & 1.6 1.9 2.4 &  \cite{Gilman2} \\
$10\div155$ ($\Delta\theta$=$10$)       & $0.5\div3.0$ ($\Delta E_{\gamma}$=$0.1$) &  \cite{Rossi} \\
\noalign{\smallskip}\hline
\end{tabular}
\end{table}
\section{ $\gamma d \rightarrow p n $ : Theoretical Overview}
\begin{figure*}
\begin{center}
\resizebox{0.7\textwidth}{!}{%
\includegraphics{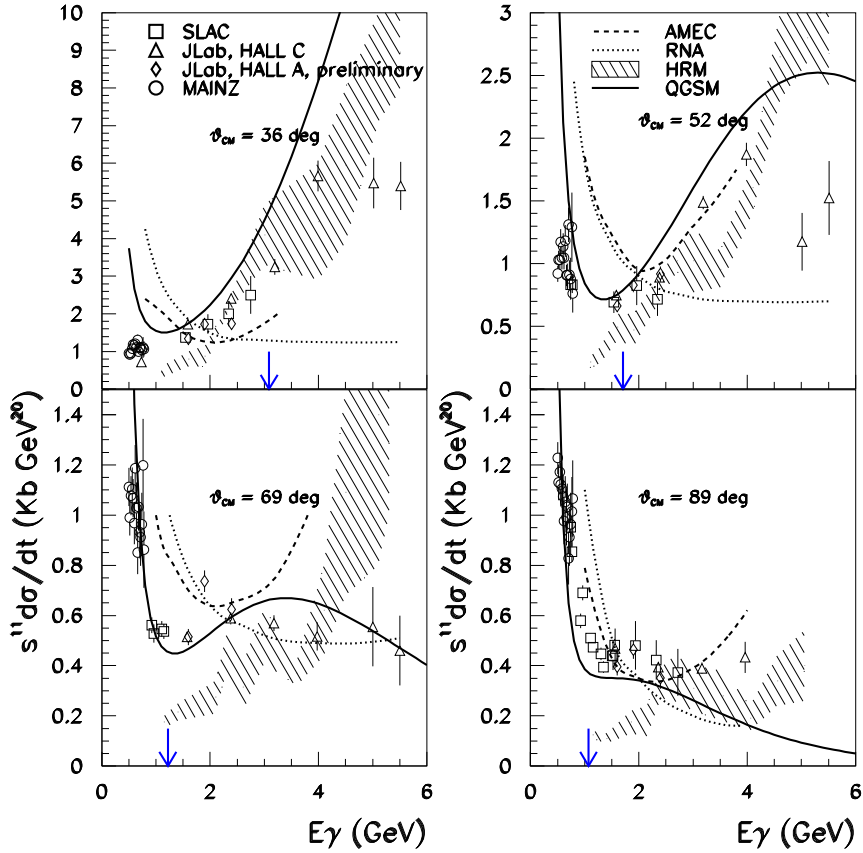}
}
\caption{Deuteron photodisintegration cross section multiplied by $s^{11}$. Experimental data are from Mainz \cite{mainz}, SLAC \cite{SLACNE81}, \cite{SLACNE82}, \cite{SLACNE17}, Hall C of TJNAF \cite{Bochna}, \cite{Schulte} and Hall A of TJNAF \cite{Gilman2}. The arrows indicate where the onset of CCR scaling is expected \cite{Matveev}, \cite{Brodsky}. For the theoretical curves, see the text.}
\label{fig:1}       
\end{center}
\end{figure*}

Predictions for the energy dependence of the cross section at high energies are given by the Constituent Counting Rules (CCR) \cite{Matveev}, \cite{Brodsky}, the Reduced Nuclear Amplitude analysis (RNA) \cite{rna}, the Hard quark Rescattering Model (HRM) \cite{hrm}, the Asymptotic Meson-Exchange Current (AMEC) \cite{amec} and the Quark-Gluon String Model (QGSM) \cite{gris}.\\
The CCR predict that, for sufficiently high energy and fixed angle, the differential cross section for any binary reaction must scale with the square of the total energy $s$:
\[\frac{d\sigma}{dt} = \frac{1}{s^{N_F - 2}} f(\theta_{CM})\]
where $N_F$ is the total number of elementary fields in the initial and final states. In our case $N_F$ = 13, thus $d\sigma/dt \propto s^{-11}$.
\\
All the available experimental data of the  $\gamma d \rightarrow p n $ differential cross section for photon energy above 1 GeV and for four proton angles, are shown in fig. \ref{fig:1}. 
In order to evidentiate the possible CCR scaling, the cross section is multiplied by $s^{11}$, and the arrows in the plots indicate the expected threshold for the onset of the scaling, $p_T > $1 GeV.
The data show scaling only at large proton angle, $\theta_{CM} = 69^{\circ}$ and $89^{\circ}$, while for more forward angles the situation is less clear. 
\\
In the RNA approch, the amplitude is described in terms of parton exchange between the two nucleons. The low energy components responsable for quark bindingin are removed by dividing out the empirical nucleon form factors and the elementary cross section is computed assuming CCR scaling.
This method gives excellent agreement with $ed$ elastic scattering \cite{rnael}, but not with the deuteron photodisintegration data for which is able to describe the cross section with an appropriate normalization factor only for $\theta_{CM} = 69^{\circ}$ and $E_{\gamma} >$ 2 GeV. This is surprising because this model is expected to approach scaling at lower energies than the Constituent Counting Rules.\\
In the HRM, it is supposed that the main contribution to the interaction comes from the absorption of the photon by a quark of a nucleon, which subsequently interacts with a quark of the other nucleon with high momentum transfer. This exchange is described by high energy, large angle, neutron-proton scattering data while the long range behavior of the deuteron is described by a deuteron wave function.
The limitation for the applicability of this model are $E_{\gamma} >$ 2.5 GeV and t $>$ 2 GeV$^2$, but, under particular assumptions for the short distance $pn$ interaction, it can be extended beyond its limits. However, the agreement with the data is poor especially at higher energies (there is a fair description of the energy dependence at $\theta_{cm} = 37^{\circ}$ and $53^{\circ}$ below $E_{\gamma} =$ 4 GeV) and the uncertainty is large, due to the poor knowledge of the $pn$ amplitude.
\\
Traditional models based on hadronic degrees of freedom are expected to fail to reproduce data at $E_{\gamma} >$ 1 GeV. 
However, this approach has been extended to the few GeV region in the Asymptotic Meson Exchange Current model, using form factors to describe the $dNN$ interaction vertex and an overall normalization factor fixed by fitting the experimental data at 1 GeV.
The results reproduce the energy dependence of the cross section only for $\theta_{cm} = 89^{\circ}$. 
It is worth noticing that also this non QCD-based model is able to provide a scaling law for the cross section, with an exponent depending on the scattering angle.
\subsection{The Quark Gluon String Model}
The Quark Gluon String Model (QGSM) is a non perturbative approach, which has been extensively used for the description of hadronic reactions at high energies \cite{kaida}. 
Due to duality property of scattering amplitudes, it can also be applied at intermediate energies for reactions without explicit resonances in the direct channel. 
In fact, it well describes the reactions $p p \rightarrow d \pi^+$ and $\bar{p} d \rightarrow M N$ \cite{qgsmadr} and has also been used for the description of heavy ions collisions at high energy \cite{ions}.
The QGSM is based on a topological expansion in QCD of the scattering amplitudes in power of $1/N$ \cite{1/n} (where $N$ is the number of colors $N_c$ or flavors $N_f$) and assumes that the scattering amplitude at high energy is dominated by the exchange of three valence quarks in the t-channel dressed by an arbitrary number of gluons.
This picture corresponds to the formation and break up of a quark-gluon string in the intermediate state, leading to the factorization of the amplitudes: the probability for the string to produce different hadrons in the final state does not depend on the type of the annihilated quarks, but is only determined by the flavour of the produced quarks.
The intermediate quark-gluon string can be easily identified with the nucleon Regge trajectory \cite{kaida}. Actually, most of the QGSM parameter can be related to the parameters of Regge Theory, as trajectories or residues. In this sense, the QGSM can be considered as a microscopic model for the Regge phenomenology, and can be used for the calculation of different quantities that have been considered before only at a phenomenological level.
\\
In Ref. \cite{gris}, the QGSM has been applied for the description of the deuteron photodisintegration reaction, using QCD motivated non linear nucleon Regge trajectories \cite{brisu}, with full inclusion of spin variables and assuming the dominance of the amplitudes that conserve s-channel helicity. 
The interference between the isoscalar and isovectorial components of the photon has been also taken into account, leading to forward-backward asymmetry in the cross section.\\
%
%
%
\begin{figure*}
\begin{center}
\resizebox{0.96\textwidth}{!}{%
\includegraphics{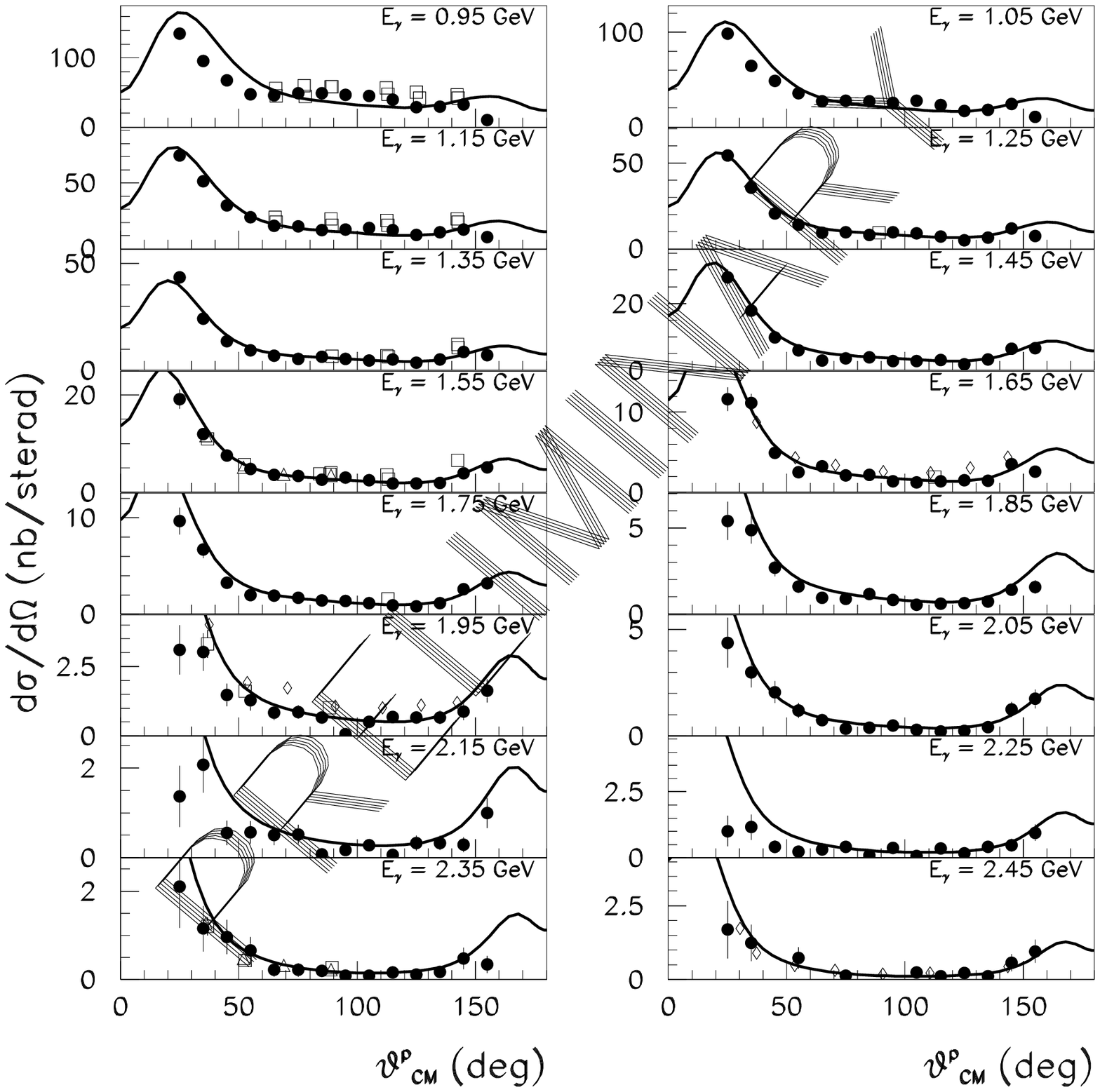}}
\vspace*{1cm}       
\caption{Preliminary results of the deuteron photodisintegration differential cross section measured in Hall B of TJNAF (black circles), compared with the published data from SLAC \cite{SLACNE81}, \cite{SLACNE82}, \cite{SLACNE17}, (open squares), Hall C of TJNAF \cite{Bochna}, \cite{Schulte} (open triangles) and preliminary data from Hall A of TJNAF \cite{Gilman2} (open diamond). The curve is the QGSM calculation \cite{gris}.}
\label{fig:2}       
\end{center}
\end{figure*}
\section{The experiment E93-017 at TJNAF}
The Experiment E93-017 was performed at TJNAF with a bremsstrahlung photon beam produced by a continuous electron beam of $E_0 =$ 2.5 and 3.1 GeV hitting a gold foil of $10^{-4}$ radiation lengths. A bremsstrahlung tagging system \cite{tagging}, with a photon energy resolution of $0.1\%$ $E_0$, was used to tag photons in the energy range from 0.5 to 3 GeV producing a photon beam of $\sim 10^7$$\gamma/s$.\\
The protons were detected in CLAS (CEBAF Large Acceptance Spectrometer)\cite{clas}, a spectrometer with nearly $4\pi$ coverage with a toroidal magnetic field generated by six superconducting coils which define six indipendent modules. Each module is equipped with 3 regions of drift chambers \cite{dc} for tracking of charged particles, and time-of-flight (TOF) scintillators \cite{tof} for charged hadron identification. 
Other detectors of CLAS (not used for the present analysis) are two electromagnetic calorimeters, covering angles up to 70 degrees, for the detection of neutral particles, and Cerenkov counters for $e/\pi$ separation.The resolution of the proton momentum is of the order of a few percent, while the proton acceptance is close to 90$\%$ in the fiducial region of the detector.\\
Real photodisintegration events were selected requiring the identification of a photon in the tagger and a proton in CLAS, and then applying a missing mass cut to the reaction $\gamma d \rightarrow p X$.\\
The CLAS acceptance and reconstruction efficiency were evaluated with Monte Carlo simulations of the photodisintegration reaction. The generated events were processed by a GEANT-based code simulating the CLAS detector, and reconstructing using the same analysis procedure that was applied to the raw data. The acceptance and reconstruction efficiency were also experimentally verified using the reaction $\gamma d \rightarrow p p \pi^{+}$.  The background contribution has been computed by a fit to the experimental missing mass distributions.
\\
The preliminary data thus far obtained for the $\gamma d \rightarrow p n $ reaction are based on about $30\%$ of the accumulated statistic and are shown in fig.\ref{fig:2}. Moreover, an additional cut $\theta_p^{CM} > 20^{\circ}$ has been applied, that will be removed in the final analysis. 
Notice that, in spite of the limited data set, the statistical error is lower than $5\%$ for $E_\gamma < 1.5$ GeV. When the whole data set will be analysed, the statistical error will be improved by a factor 3 or more for photon energies up to 2.5 GeV.
The particular striking and unique feature of the present data is the broad angular and energy range measured with the CLAS detector. This allows to highlight a persistent forward backward-asymmentry in the differential cross section all over the energy range. The data also show good agreement with previous measurement \cite{SLACNE81},\cite{SLACNE82},\cite{SLACNE17},\cite{Bochna},\cite{Schulte},\cite{Gilman2} in the overlapping regions.
In the figure is also shown as full line the results of the QGSM calculation. The agreement between data and calculations is very good for all energies: the QGSM reproduces the angular dependance of the cross section at fixed photon energy and predicts the forward-backward asymmetry showed by the data.
From these plots, it results also the importance of measuring the cross section at very forward angles, in order to check the QGSM prediction of a decrease of the cross section for angles smaller than $10^{\circ}-20^{\circ}$.\\
\begin{figure*}
\begin{center}
\resizebox{0.7\textwidth}{!}{%
\includegraphics{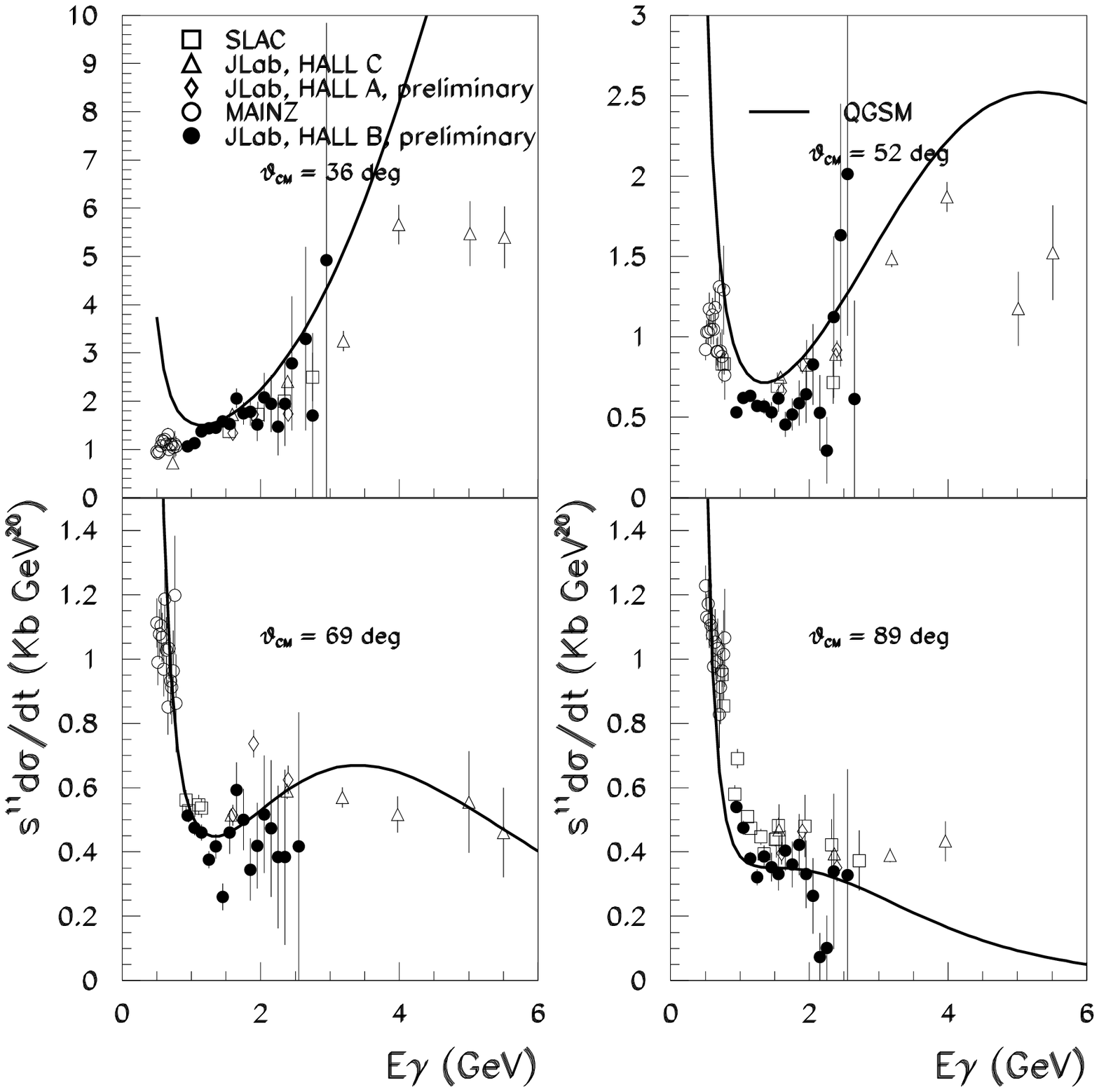}
}
\caption{Deuteron photodisintegration cross section multiplied by $s^{11}$ for four fixed proton angles: preliminary results of Hall B of TJNAF (black circles) and published data from Mainz \cite{mainz}, SLAC \cite{SLACNE81}, \cite{SLACNE82}, \cite{SLACNE17}, Hall C of TJNAF \cite{Bochna}, \cite{Schulte} and Hall A of TJNAF \cite{Gilman2}. The curve is the QGSM calculation \cite{gris}.}
\label{fig:3}       
\end{center}
\end{figure*}
The QGSM also predicts correctly the decrease of $d\sigma / dt$ at fixed angle as a function of the photon energy, as shown in fig. \ref{fig:3} for four CM proton angles. Also here, the cross section is multiplied by the factor $s^{11}$.
\section{Conclusions}
The study of high-energy two-body photodisintegration of the deuteron has received a renewed interest in recent years with attempts to illuminate the quark structure of the nucleus.
The differential cross section of this reaction has been measured for the first time with nearly compleate angular coverage in the energy range between 0.5 and 3 GeV in the Hall B of Jlab. The preliminary results of about of $30\%$ of the total statistic accumulated and photons with energy between 0.9 and 2.5 GeV show persistent forward-backward asymmetry and are in good agreement with the few published data in the overlapping region.
The QGSM well reproduces the data for all proton angles and photon energy above 1 GeV. In the final analysis the points at $10^{\circ}-20^{\circ}$ will allow to check the decrease of the cross section predicted by the QGSM.


\begin{thebibliography}{}
\bibitem{holt}
R. J. Holt, Phys. Rev. C \textbf{41}, (1990) 2400. 
\bibitem{SLACNE81}
J. Napolitano et al., Phys. Rev. Lett. \textbf{61}, (1988) 2530.
\bibitem{SLACNE82}
S. J. Freedman et al., Phys. Rev. C \textbf{48}, (1993) 1864.
\bibitem{SLACNE17}
J. E. Beltz et al., Phys. Rev. Lett. \textbf{74}, (1995) 646.
\bibitem{Bochna}
C. Bochna et al., Phys. Rev. Lett. \textbf{81}, (1998) 4576.
\bibitem{Schulte}
E. C. Schulte et al., Phys. Rev. Lett. \textbf{87}, (2001) 102302-1.
\bibitem{Gilman1}
K. Wijesooriya et al., Phys. Rev. Lett. \textbf{86}, (2001) 2975.
\bibitem{Gilman2}
R. Gilman et al., CEBAF Experiment \textbf{E99-008}, (1999) .
\bibitem{Rossi}
P. Rossi et al., CEBAF Experiment \textbf{E93-017}, (1993) .
\bibitem{Matveev}
V.A. Matveev, R.M. Muradyan and A.N. Tavkhelidze, Lett. Nuovo Cimento \textbf{7}, (1973) 719
\bibitem{Brodsky}
S. J. Brodsky and G. Farrar, Phys. Rev. Lett. \textbf{31}, (1973) 1153
\bibitem{rna}
S. L. Brodsky and J. R. Hillier, Phys. Rev. C, \textbf{28}, (1983) 475.
\bibitem{hrm}
L. L. Frankfurt et al.,  Phys. Rev. Lett. \textbf{84}, (2000) 3045;\\
L. L. Frankfurt et al.,  Nucl. Phys. A \textbf{663\&664}, (2000) 34.
\bibitem{amec}
A.E.L. Dieperink and S. I. Nagorny, Phys. Lett. B \textbf{456}, (1999) 9.
\bibitem{gris}
V. Yu. Grishina et al.,  Eur. Phys. J. \textbf{A10}, (2001) 35.
\bibitem{rnael} 
S. J. Brodsky and B. Chertok, Phys. Rev. D \textbf{14}, (1976) 3003.
\bibitem{kaida} 
A. B. Kaidalov, Z. Phys. C \textbf{12}, (1982) 63;\\
A. B. Kaidalov, Surv. High Energy Phys. \textbf{13}, (1999) 265.
\bibitem{qgsmadr}
A. B. Kaidalov, Sov. J. Nucl. Phys. \textbf{53}, (1991) 872;\\
C. Guaraldo et al., Yad. Fiz. \textbf{59}, (2000) 1896;\\
C. Guaraldo et al., Phys. Atom. Nucl. \textbf{63}, (2000) 1395.
\bibitem{ions} 
E. E. Zabrodin et al., Phys. Lett. B \textbf{508}, (2001) 184.
\bibitem{1/n} 
G. t'Hooft, Nucl. Phys. B \textbf{72}, (1974) 461.
\bibitem{brisu} 
M. M. Brisudova et al., Phys. Rev. D \textbf{61}, (2000) 054013.
\bibitem{mainz}
R. Crawford et al., Nucl. Phys. A \textbf{603}, (1996) 303.
\bibitem{tagging}
D. I. Sober et al., Nucl. Inst. and Meth. A \textbf{440}, (2000) 263.
\bibitem{clas}
W. Brooks, Nucl. Phys. A \textbf{663\&664}, (2000) 1077c.
\bibitem{dc}
M. D. Mestayer et al., Nucl. Inst. and Meth. A  \textbf{449}, (2000) 81.
\bibitem{tof}
E. S. Smith et al., Nucl. Inst. and Meth. A \textbf{432}, (2000) 265.

\end{thebibliography}
\end{document}